\documentclass[%
 reprint,
nobibnotes,
bibnotes,
 amsmath,amssymb,
 aps,
]{revtex4-1}

\usepackage{multirow}
\usepackage{amsmath}
\usepackage{booktabs}
\usepackage{graphicx}
\usepackage{dcolumn}
\usepackage{bm}

\begin{document}

 \preprint{APS/123-QED}

\title{Direct Counting Analysis on Network Generated by Discrete Dynamics}

\author{Shu-Hao Liou}
 \email{hellont92@gmail.com}
\author{Chia-Chu Chen}
 \email{chiachu@phys.ncku.edu.tw}
 \affiliation{Department of Physics, National Cheng Kung University, Tainan, Taiwan 70101}
\date{\today}

\begin{abstract}
A detail study on the In-degree Distribution (ID) of Cellular
Automata is obtained by exact enumeration. The results indicate
large deviation from multiscaling and classification according to ID
are discussed. We further augment the transfer matrix as such the
distributions for more complicated rules are obtained. Dependence of
In-degree Distribution on the lattice size have also been found for
some rules including $\textbf{R}50$ and $\textbf{R}77$.
\begin{description}
\item[PACS numbers]
89.75.Hc, 89.75.Da, 89.75.Fb.
\end{description}
\end{abstract}

\maketitle

Discrete dynamical system in general can generate very complicated
dynamics even though the rule or the equation of evolution for the
system can be apparently simple. Examples include the
synchronization of Pulse-coupled Oscillators, chaos arises from
discrete mapping and more interestingly the simple Cellular Automata
(CA) introduced by von Neumann almost half a century ago
\cite{neuman66}. One of the important problems in discrete dynamical
system is the classification of it's general behavior. Generally
speaking this is a hard question to be answered since the
classification requests a full understanding of the dynamics itself.
Even for simple system such as CA which evolves according to simple
rules, a complete classification is still controversial since the
work of StevenWolfram in the 80' of last century \cite{wolfram84}.
Recently, A. Shreim \textit{et al}. have approached the CA problem
by employing concepts originated from network analysis
\cite{shreim07}. They classified CA by analyzing the local and
global properties of the network generated by the dynamics of CA.
Their approach certainly provides new way to address this
interesting and difficult problem. In this work we follow up their
approach and investigate further on this problem. Here we would only
concentrate on the local analysis and will return on the global
property later.

For any discrete dynamical system, the configuration space can be
represented by discrete nodes. The time evolution of the system in
each time step is presented by a connected link starting from one
node to its dynamical successor. As a result, all the trajectories
of the dynamical system generated a directed network. Since CA are
deterministic systems, each node possess one single outgoing link.
However, due to the existence of fixed points and periodic
solutions, the dynamics of CA in general is irreversible and the
number of preimages of a state can be larger than $1$. The in-degree
(ID) of a node is defined as it's number of preimages and obviously
in CA the in-degree is a local property. In \cite{shreim07} the
authors have performed analysis on the ID distribution for various
one dimensional CA both analytically and numerically. Due to the
simplicity of Rule $4$ ($\textbf{R}4$), by using a transfer matrix
(\textbf{T}) approach together with combinatory analysis on
preimages, they obtained an approximation solution of the
probability distribution $P(k)$ with $k$ denoting the in-degree.
Their results were presented on the $\log-\log$ plot as curves
instead of straight lines and seem to suggest the existence of
multiscaling distribution. In our work, instead of using
approximation, we re-analyze this problem by performing direct
counting of the in-degree for various rules with different lattice
size $L$. What we have found is that the distribution is not as
simple as what have been obtained in \cite{shreim07}. The plan of
this report is as follows. A brief summarization on nomenclatures of
CA and network are given and the results on ID of \cite{shreim07}
are briefly discussed in section \ref{sec:level1}. The direct
counting results of various rules are reported in section
\ref{sec:level2} where some further studies on the ID are also
treated. The extension of transfer matrix for more accurate analysis
on other rules and its implication are treated in Section
\ref{sec:level3}. A brief summary of our findings is provided in the
final section.

\section{\label{sec:level1}INTRODUCTION}

A one-dimensional Cellular Automaton is defined on a spatial lattice
of $L$ sites where $L$ can either be finite or infinite. In this
work we will keep $L$ finite and the state of each site is in one of
the $g$ states at any time $t$. Each site follows the same
prescribed rules for updating. For elementary CA, the number of
neighborhood of each site is $2$. The state of CA starts out with
arbitrary initial configuration which is represented by
$S(0)=\{s_0,s_1,s_2,\ldots,s_L\}$ where $s_i$ can be in any one of
the $g$ states. The configuration of system at time $t$ is denoted
by $S(t)=\{s^t_0,s^t_1,s^t_2,\ldots,s^t_L\}$. In this work the state
of each site is restricted to $g=2$ such that $s_i\in\{0,1\}$. Let
$\textbf{R}$ be any one of the CA, the value of $s_i$ at $t+1$ time
step is set equal to $R(s^t_{i-1}s^t_{i}s^t_{i+1})$ which is equal
to $0$ or $1$ according to the rule of interest. Apparently for
elementary CA there are $256$ rules. Following Wolfram's notation
each rule can be assigned with a number given by:
\begin{eqnarray*}
R(000)+2R(001)+2^2R(010)+2^3R(011)+2^4R(100)\\
+2^5R(101)+2^6R(110)+2^7R(111).
\end{eqnarray*}
For example, Rule $237$ ($\textbf{R}237$) is the one with
$R(100)=0$, $R(001)=0$ and all others $R(s_1s_2s_3)=1$. Based on
large number of numerical studies, Wolfram has suggested that CA can
be classified into four classes \cite{wolfram84}. Cellular Automata
within each class have the same qualitative behavior. Starting from
almost all initial conditions, trajectories of CA become
concentrated onto attractors, the four classes can then be
characterized by their attractors. According to Wolfram's
classification, classes I, II and III are roughly corresponding to
the limit points, limit cycles and chaotic attractors in continuous
dynamical systems respectively. More precisely their respective long
time limits are: (I) spatially homogeneous state, (II) fixed
(steady) or periodic structure and (III) chaotic pattern throughout
space. The fourth class of CA behaves in a much more complicated
manner and was conjectured by Wolfram as capable for performing
universal computation. For finite one-dimensional CA, the lattice is
arranged on a circle with periodic boundary conditions. Such
Cellular Automata have a finite number of states $N=2^L$, and as a
result, after sufficient long time evolution the system must enter
the state which can either be homogenous, steady or periodic state.
Therefore, the class III and IV attractors do not exist in finite
CA. However, it is also known that classes III and IV are hard to
distinguished in some cases and resulting with disputed
classification. In some sense this is due to the lack of a sensible
way for defining complexity which is the criterion for class IV. In
\cite{shreim07} A. Shreim \textit{et al}. studied networks generated
by CA, they claimed to have found highly heterogeneous state space
networks for classes III and IV in contrast to the networks
generated by classes I and II. One of the characteristic of state
space is the in-degree distribution $P(k)$ which is defined as
\cite{shreim07}
\begin{equation}
P(k)=\frac{\Sigma}{N}. \label{eq:one}
\end{equation}
where $L$ is the size of the lattice, $N=2^L$ is the total number of
state and $\Sigma$ is the number of state with in-degree $k$.
However in their analysis on the local properties, approximation has
been applied in the analytical calculation of ID which shows
nontrivial scaling. For completeness we briefly summarize their
method of calculation in this section. They introduced the transfer
matrix $\textbf{T}$ which maps each pair $s_{i-1}s_i$ onto the pair
$s_is_{i+1}$, and such $\textbf{T}^{(s)}$ could be used to
characterize all the preimages of hubstates (A hubstate is the state
which has the maximum in-degree $k_m$.) The rows and columns are
order as ``$00$'', ``$01$'', ``$10$'' and ``$11$'' accordingly. For
example the $\textbf{T}$ matrix of $\textbf{R}4$ is given by:
\begin{eqnarray}
\textbf{T}_4 = \left(
\begin{array}{cccc}
1 &1 &0 &0\\
0 &0 &0 &1\\
1 &1 &0 &0\\
0 &0 &1 &1
\end{array}\right).\label{eq:two}
\end{eqnarray}
It is easy to check that the maximum in-degree $k_m$ is related to
$\textbf{T}$ as
\begin{equation}
k_m=Tr(\textbf{T}^L). \label{eq:three}
\end{equation}
The scaling of $k_m$ with the lattice size can be obtained by using
the largest eigenvalue $\lambda_m$ of $T$, namely,
$k_m\approx{N^\nu}$ with $\nu=\log_2{\lambda_m}$. Furthermore, for
$\textbf{R}4$ we assigned $n$ as the number of isolated $1's$, for
example two configurations $(010000)$ and $(010100)$ have $n=1$ and
$2$ respectively. Then by assuming a one-one corresponding relation
between $k$ and $n$, the following expression was proposed for the
in-degree distribution:
\begin{equation}
P(k)dk=\frac{\Omega(n)dn}{2^L}. \label{eq:four}
\end{equation}
A multiscaling result is then obtained and given by the following
expression \cite{shreim07}:
\begin{equation}
y=-1-x+\log_2{\left[\frac{(1-\epsilon)^{1-\epsilon}}{\epsilon^{\epsilon}(1-2\epsilon)^{1-2\epsilon}}\right]}.
\label{eq:five}
\end{equation}
where $y=\log{P(k)}/{\log{N}}$, $x=\log{k}/{\log{N}}$ and
$\epsilon=n/L$. The result of Eq. (\ref{eq:five}) was ploted in
\cite{shreim07} and are apparent curvature appeared on the curve as
such a conclusion of nultiscaling method than finite-size effect was
proposed. Similar results of multiscaling in other rules were also
reported in their article by using the same approximation approach.
To compare the results of \cite{shreim07} with exact calculation we
have performed the exact enumeration of $\textbf{R}4$ with different
$L$. For the lack of space, in Table \ref{tab:table1} we only list
all the in-degree's for $L=12$. The same characteristic also exist
for other values of $L$. One notices that for the same $k$ there
corresponds more than one value of $n$. For example, for $k=12$, $n$
can either be $2$ or $3$. Therefore the meaning of $dn/dk$ is
ambiguous. One might argue that the multi-valueness of $n$ is an
artifact of small $L$. However we have checked with $L=10^4$ where
different states with the same $k$ were found. In fact, for $k=1$,
states of different $n$ can be constructed easily. For example, for
$k=1$, we have found at least $2$ states $(000100010001\ldots)$ and
$(010101\ldots)$ which correspond to $n=2500$ and $5000$
respectively. Since the results were obtained for small $k$, there
is a possibility that dn/dk might be well-defined for large $k$. To
refute this reasoning, we have performed the large $k$ analysis for
different $L$. For $L=12$, $k_m=852$ with $n=0$ and the second
maximum in-degree $k_{2m}=114$ with $n=1$; even in $L=25$, we got
$k_m=1276941$ with $n=0$ and $k_{2m}=170625$ with $n=1$. The results
indicate that for $\triangle{n}=1$ the $\triangle{k}=k_m-k_{2m}$ is
not a small quantity. In fact $\triangle{k}$ rises exponentially as
$L$ getting larger. As a result, $dk$ can never be infinitesimal
implying the meaningless of $dn/dk$. Thus the simple multiscaling
results in \cite{shreim07} is in doubt.

\begin{figure}
\includegraphics{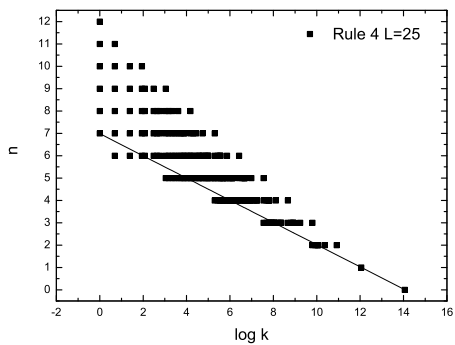}
\caption{\label{fig:one} The relation of $n$ and in-degree $k$ for
Rule $4$. The solid curve represents Eq. (6) in \cite{shreim07}.}
\end{figure}

The result of Eq. (\ref{eq:four}) is presented by the solid line in
Figure \ref{fig:one}. By substituting the corresponding $k_m$ in Eq.
(\ref{eq:five}), one has $y=-1.8$. For $\textbf{R}4$ with $L=25$,
the results of $n-k$ relation are provided in Figure \ref{fig:one}.
There is only one hubstate with $n=0$ in such system, consequently
the exact $P(k_m)$ equals to ${1}/{2^L}$ \cite{newman03} which
implies $y=-1$ independent of $L$ instead of $-1.8$ as obtained by
the approximation calculation. Obviously the approximation for
getting Eq. (\ref{eq:five}) is inadequate. In fact the above
discrepancies prompted us to re-analyze this problem.
\begin{table}
\caption{\label{tab:table1} The relation of in-degree $k$ and $n$
for $\textbf{R}4$ with $L=12$. }
\begin{ruledtabular}
\begin{tabular}{ccc|ccc}
 & &\textrm{number of}& & &\textrm{number of}\\
\textrm{$k$}& \textrm{$n$}&\textrm{states}&
\textrm{$k$}& \textrm{$n$}&\textrm{states}\\
\colrule
    852 & 0 & 1 & 7 & 3 & 24 \\
    114 & 1 & 12 & 4 & 3 & 48 \\
    37 & 2 & 12 &  & 4 & 12 \\
    21 & 2 & 12 & 2 & 3 & 24 \\
    16 & 2 & 6 &  & 4 & 3 \\
    14 & 2 & 12 & 1 & 3 & 4 \\
    14 & 2 & 12 &  & 4 & 57 \\
    12 & 2 & 12 &  & 5 & 36 \\
    & 3 & 12 &  & 6 & 2 \\
\end{tabular}
\end{ruledtabular}
\end{table}
\section{\label{sec:level2}NETWORK CHARACTERISTICS BY
DIRECT ENUMERATION} As suggested in \cite{shreim07}, local
properties such as the scaling of maximum in-degree with lattice
size and the in-degree distribution might be useful for
characterizing networks and hopefully also provide some help to
classify CA. However, from previous discussions it is clear that
approximation calculations can lead to inadequate conclusions such
as multiscaling of ID being an example. In order to have a better
understanding on this kind of network, it is unavoidable to address
the problem with exact enumeration which will be the approach
adopted in this work. We have studied more rules from different
classes suggested by Wolfram \cite{wolfram84} and the in-degree
distributions are evaluated exaclty by direct counting method. The
transfer matrix approach proposed in \cite{shreim07} has been used
in this work extensively.

It is important to note at the outset that not all rules can have a
transfer matrix representation. The following matrix represents
possible candidate of $\textbf{T}$ for elementary CA:
\begin{eqnarray}
\textbf{T} = \left(
\begin{array}{cccc}
a_1 & a_2 &0 &0\\
0 &0 &a_3 &a_4\\
a_5 &a_6 &0 &0\\
0 &0 &a_7 &a_8
\end{array}\right).\label{eq:six}
\end{eqnarray}
where $a_i$, $i=1\sim8$, can take on either $1$ or $0$. In fact
there are totally $256$ different matrices represented by
$\textbf{T}$, however it is also known that not all of them can
represent the $256$ elementary CA. This is partly due to the fact
that not all rules can have a $\textbf{T}$ matrix representation. In
particular, by diagonalizing all possible $\textbf{T}$, we have
obtained only $9$ different $\lambda_m$: $1$, $1.22074$, $1.32472$,
$1.38028$, $1.46557$, $1.61803$, $1.75488$, $1.83929$ and $2$. Some
of the rules corresponding to different $\lambda_m$ are listed in
Table \ref{tab:table3}.

\begin{table}
\caption{\label{tab:table3} The classification for different rules
with $\lambda_m$. }
\begin{ruledtabular}
\begin{tabular}{lll}
 \textrm{$\lambda_m$}&\textrm{$\nu$}&\textrm{Rules}\\
 \colrule
 1.32472& 0.4056& 110,62\\
1.46557 &0.551& 232,33,6\\
 1.61803 &0.6942 &36,18,5\\
  1.75488 &0.811 &4,32\\
 2& 1 &0,255\\
\end{tabular}
\end{ruledtabular}
\end{table}

\begin{figure}
\includegraphics[width=0.5\textwidth]{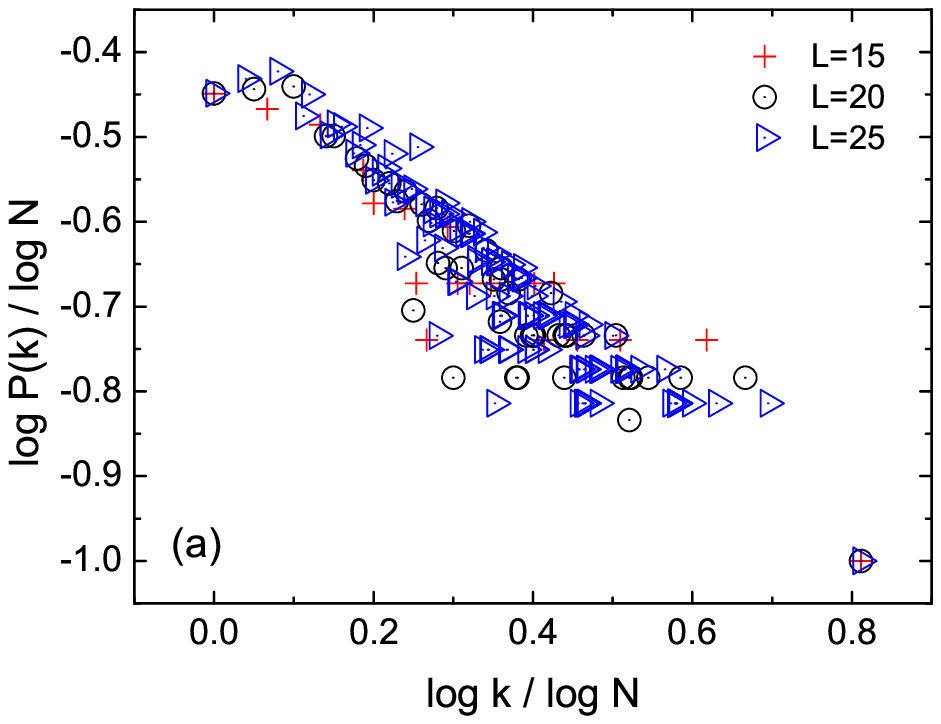}\\
\includegraphics[width=0.475\textwidth]{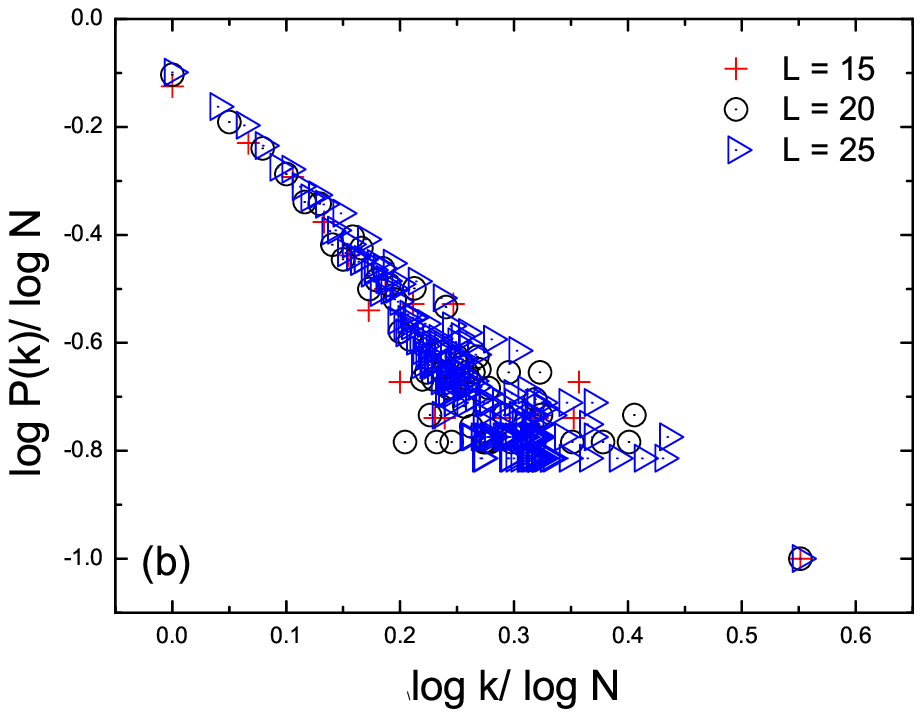}\\
\includegraphics[width=0.5\textwidth]{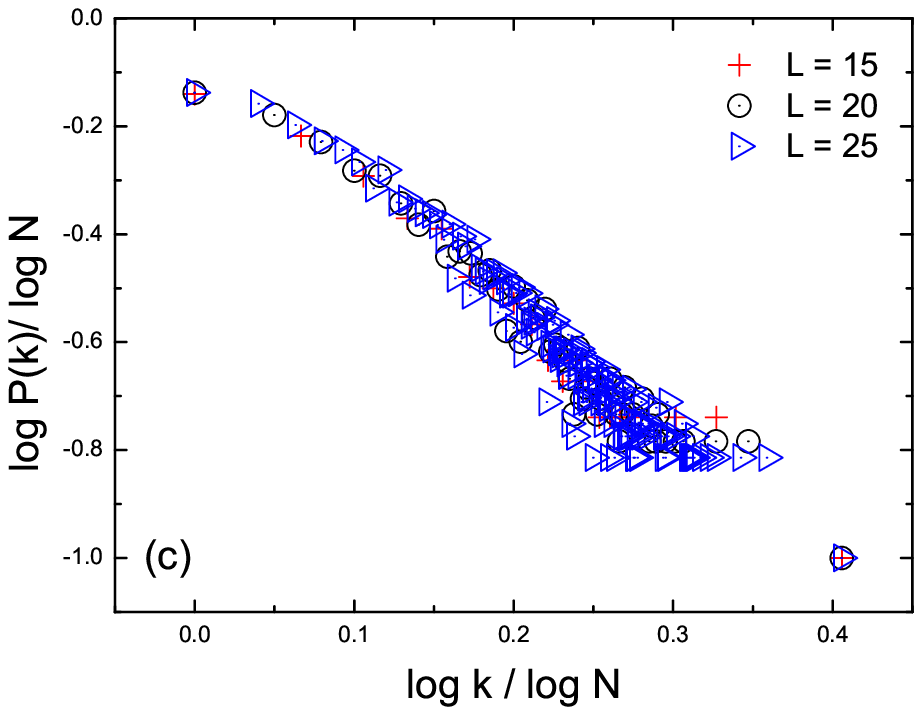}\\
\caption{\label{fig:two} The direct counting results of ID for
(a)$\textbf{R}4$, (b)$\textbf{R}22$ and (c)$\textbf{R}110$ with
different $L$.}
\end{figure}

\begin{figure}
\includegraphics{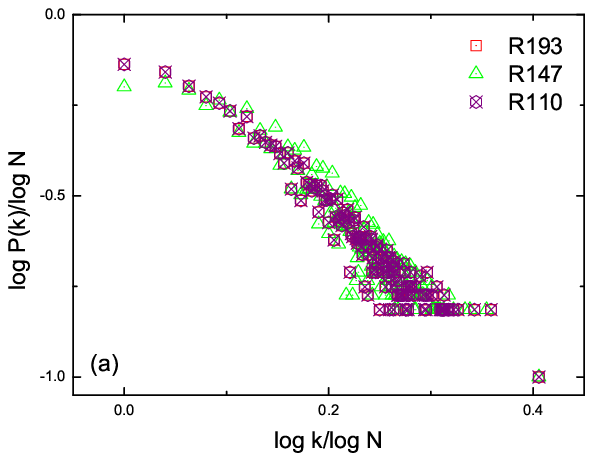}\\
\includegraphics{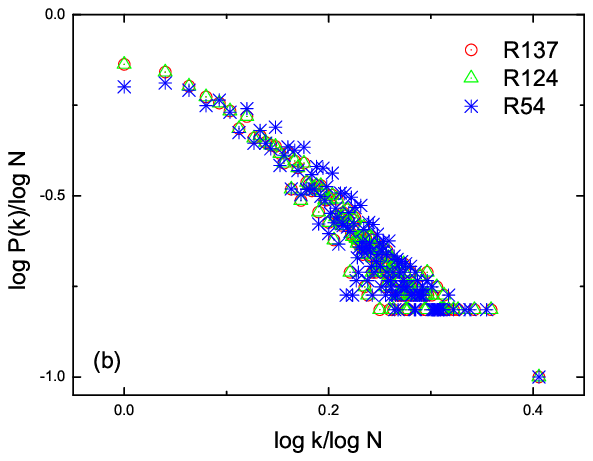}\\
\caption{\label{fig:three} The In-degree Distributions for (a)
$\textbf{R}193$, $\textbf{R}147$, $\textbf{R}110$ and (b)
$\textbf{R}137$, $\textbf{R}124$ and $\textbf{R}54$ with $L=25$.}
\end{figure}

\begin{figure}[h]
\includegraphics{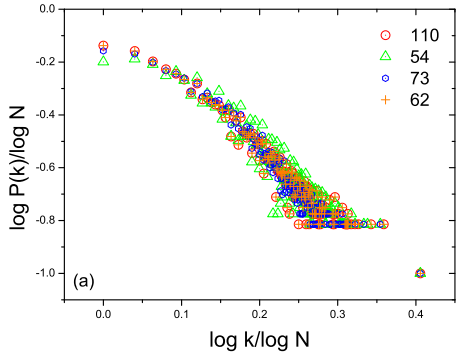}\\
\caption{\label{fig:four} The In-degree Distributions for
$\textbf{R}110$, $\textbf{R}54$, $\textbf{R}73$ and $\textbf{R}62$
with $L=25$.}
\end{figure}

\begin{figure}
\includegraphics{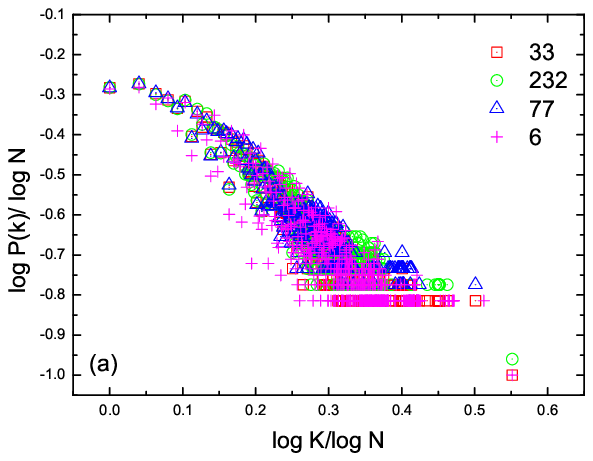}\\
\includegraphics{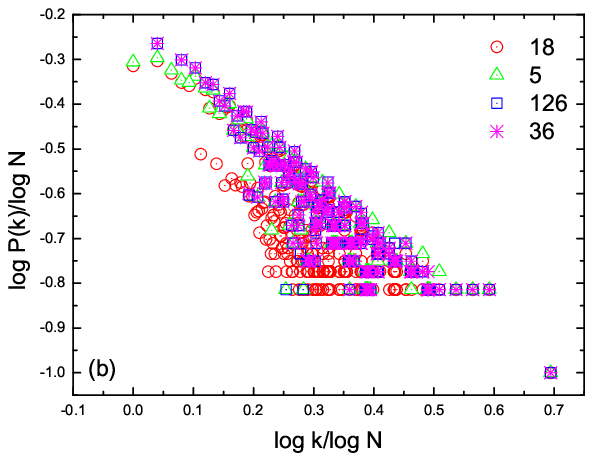}\\
\includegraphics{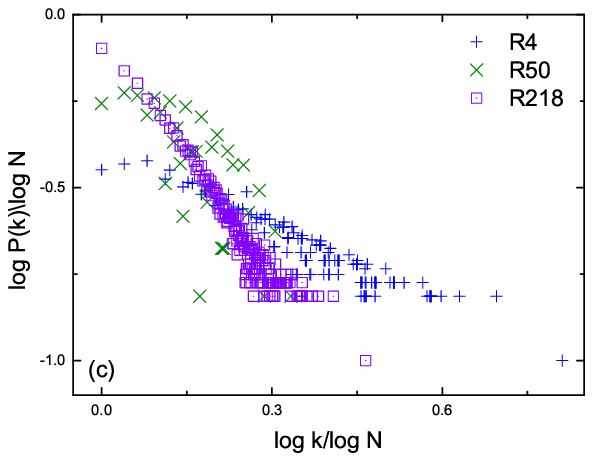}
\caption{\label{fig:five} The
in-degree distributions for various rules with (a) $\nu=0.551$ and
(b) $\nu=0.6942$; (c) The in-degree distributions for Class II,
where $\nu=0.811$ for $\textbf{R}4$, $\nu=0.3471$ for $\textbf{R}50$
and $\nu=0.4650$ for $\textbf{R}218$. All of results are ploted with $L=25$.}
\end{figure}

In what follows, all ID are evaluated by direct counting. To set the
stage we present the in-degree distributions of various rules with
different $L$ in Fig. \ref{fig:two}, the results of $\textbf{R}4$,
$\textbf{R}22$ and $\textbf{R}110$ are plotted. The results of each
rule seem to show the same distribution for three different values
of $L$. Especially for small $k$ the distributions are in good
agreement for different $L$ with all the points fall on a simple
curve with non-vanishing curvature. As $k$ increases, the
distribution becomes more spread out but still maintains the same
distribution for various $L$. These observation seems to suggest
that the ID of this kind of network is scale independent. Moreover,
since these rules belongs to different classes and the scale
independent of the distributions seem to be universal for all four
classes of CA. Similar results can also be obtained for other rules
and will not be shown explicitly. Even though the $L$ independence
of ID is very suggestive, however a definite answer would require a
full scale analysis with larger lattice size. Work along this line
is now being studied and will be reported separately. On a different
front, with knowing all the possible values of $\lambda_m$, it is
natural to wonder if rules with the same $\lambda_m$ are correlated.
To show whether such correlation exists or not, results of the same
$\nu=\log_2\lambda_m$ are plotted in Fig. \ref{fig:three} and
\ref{fig:four}. For $\nu=0.4056$, the ID of $\textbf{R}193$,
$\textbf{R}147$, $\textbf{R}110$, $\textbf{R}137$, $\textbf{R}124$
and $\textbf{R}54$ are given in Fig. \ref{fig:three}, all of which
show similar distribution. Interestingly these rules belong to the
same class, namely class four according to Wolfram's scheme. For
clarity, in Fig. \ref{fig:four} we have also plotted the ID of other
rules of class IV. They also share the same feature as given in Fig.
\ref{fig:three}. Similarly, the result for other are given in Fig.
\ref{fig:five}(a) with $\nu=0.551$, the corresponding rules are all
of class II \cite{kayama95} and the ID of each rule also shows a
universal distribution. However such good correlation between ID and
$\nu$ is not conclusive. Furthermore rules from different classes
can have the same $\nu$. For example, the plot given in Fig.
\ref{fig:five}(b) for the in-degree distributions of various rules
corresponding to $\nu=0.6942$ shows diverse patterns. It is noted
that in this plot, $\textbf{R}36$ and $\textbf{R}5$ are class II,
whereas $\textbf{R}18$ and $\textbf{R}126$ are classified as class
III \cite{kayama95}. To make things worse, the plot for various
rules of class II is given in Fig. \ref{fig:five}(c), where
apparently the patterns get more diverse. More importantly, those
rules belong to different $\nu$'s. As a result, one might conclude
that there are good correlation for class IV which possess
$\nu=0.4056$, apart from that the ID of other classes do not share a
universal pattern and also the value of $\nu$ can not be used as a
indicator for classification. Thus the local properties such as
$\nu$ and in-degree distribution are not useful for CA
classification as suggested in \cite{shreim07}.
\section{\label{sec:level3}AUGMENTATION OF T MATRIX}
\begin{figure}
\includegraphics{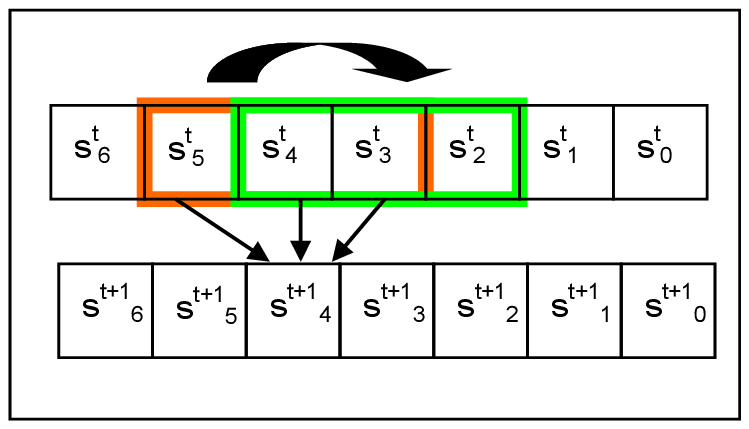}\\
\caption{\label{fig:six} The diagram for $8\times8$
$\widetilde{\textbf{T}}$ matrix.}
\end{figure}

The $\textbf{T}$ matrix approach for calculating maximum in-degree
is quite efficient and there are also cases in which the in-degree
can also be evaluated exactly. However, as clearly stated in the
previous section that $4\times4$ $\textbf{T}$ matrix do not always
exist for elementary CA and therefore it is interesting to see how
modification on the transfer natrix scheme can be extended to other
rules. In this section an extension is suggested by augmenting the
$4\times4$ matrix to a $8\times8$ matrix. Instead of mapping each
pair $s^t_{i-1}s^t_i$ onto the pair $s^t_{i-1}s^t_i$,
$\widetilde{\textbf{T}}$ will map $s^t_{i-2}s^t_{i-1}s^t_i$ into the
triplet $s^t_{i-1}s^t_{i}s^t_{i+1}$ and hence define a $8\times8$
$\widetilde{\textbf{T}}$ matrix which could be used to describe the
preimage more efficiently. The basis states are
\begin{equation*}
\mid{i}\rangle\in{\{(111),(110),(101),(100),(011),(010),(001),(000)\}}.
\end{equation*}
Fig. \ref{fig:six} shows the definition of $8\times8$
$\widetilde{\textbf{T}}$ matrix.

There are some merits in this approach. First of all, for elementary
CA where all the rules of evolution involve three adjacent cells,
obviously such extension certainly respect the evolution dynamics
and hence include more information of the dynamical system.
Secondly, the augmentation of $\textbf{T}$ indeed resolves some of
the cases which can not be done with $4\times4$ matrices. Moreover,
this approach can also be applied to other more complicate CA
including the model with next nearest-neighbor interaction, we will
illustrate that by an example in this section.

\begin{figure}
\includegraphics[width=0.5\textwidth]{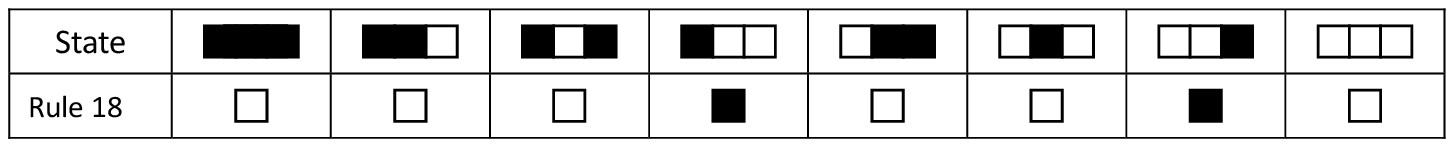}\\
\caption{\label{fig:seven} The diagram of $\textbf{R}18$.}
\end{figure}

It is easy to see that whenever the $4\times4$ transfer matrix can
be defined then there is always a corresponding $8\times8$ matrix
however the reverse is not always true. For the case when both
descriptions exist then a natural question about the uniqueness of
the results arises, namely, can both $4\times4$ and $8\times8$
matrices provide the same results? This is due to the fact that the
eigenvalues of both matrices can be very much different and there is
no obvious reason for them to be the same. For all the cases we have
studied the maximum in-degree can be exactly calculated by both
$4\times4$ and $8\times8$ matrices and the results are identical.
Unfortunately there exists no analytical proof. To illustrate this
fact, the following example is the calculation of $k_m$ with the
$\widetilde{\textbf{T}}$ matrices for $\textbf{R}_{18}$. Fig.
\ref{fig:seven} is the evolution rule of $\textbf{R}_{18}$, and the
$4\times4$ $\textbf{T}$ matrix obtained in Amer Shreim \textit{et
al.}'s study \cite{shreim07} is:
\begin{eqnarray}
\textbf{T}_{18}=\left(
\begin{array}{cccc}
1 & 0 &0 &0\\
0 &0 &1 &1\\
0 &1 &0 &0\\
0 &0 &1 &1
\end{array}\right).\label{eq:seven}
\end{eqnarray}
The hubstate of $\textbf{R}18$ is $(00\dots00)$, then Fig
\ref{fig:seven} shows that $(001)$ and $(100)$ are excluded from the
pre-images of $(00\dots00)$, therefore one has
$\widetilde{\textbf{T}}_{18}$:
\begin{eqnarray}
\widetilde{\textbf{T}}_{18} = \left(
\begin{array}{cccccccc}
1 & 0 &0 &0& 0 &0 &0 &0\\
0 &0 & 0 &0 &0& 0 &0 &0\\
0 &0 & 0 &0 &0& 1 &0 &0\\
0 &0 & 0 &0 &0& 0 &1 &1\\
0 &0 & 0 &0 &0& 0 &0 &0\\
0 &0 & 1 &1 &0& 0 &0 &0\\
0 &0 & 0 &0 &0& 1 &0 &0\\
0 &0 & 0 &0 &0& 0 &1 &1
\end{array}\right).\label{eq:eight}
\end{eqnarray}

The in-degree of hubstate is given by $k(S)=Tr(T^L)$, with $L=10$
one has:
\begin{eqnarray}
\textbf{T}_{18}^{10} = \left(
\begin{array}{cccc}
1 & 0 &0 &0\\
0 &34 &55 &55\\
0 &21 &34 &34\\
0 &34 &55 &55
\end{array}\right).\label{eq:nine}
\end{eqnarray}
and
\begin{eqnarray}
\widetilde{\textbf{T}}_{18}^{10} = \left(
\begin{array}{cccccccc}
1 & 0 &0 &0& 0 &0 &0 &0\\
0 &0 & 0 &0 &0& 0 &0 &0\\
0 &0 & 13 &13 &0& 21 &21 &21\\
0 &0 & 21 &21 &0& 34 &34 &34\\
0 &0 & 0 &0 &0& 0 &0 &0\\
0 &0 & 21 &21 &0& 34 &34 &34\\
0 &0 & 13 &13 &0& 21 &21 &21\\
0 &0 & 21 &21 &0& 34 &34 &34
\end{array}\right).\label{eq:ten}
\end{eqnarray}
It is clear that the traces of these two matrices are the same and
resulting the same value for the maximum in-degree $k_m$.

We have also obtained the $8\times8$ $\widetilde{\textbf{T}}$ matrix
for more than $22$ rules and also rules related to them by symmetry(
by exchanging $0\leftrightarrow{1}$ ). They are: $1$, $2$, $4$, $5$,
$6$, $7$, $18$, $19$, $22$, $23$, $33$, $36$, $40$, $54$, $62$,
$73$, $94$, $110$, $160$, $218$ and $232$. Except for
$\textbf{R}94$, all these rules possess both $\textbf{T}$ and
$\widetilde{\textbf{T}}$ which results with the same $k_m$. Further
application of this augmentation procedure is illustrated by the
example of the majority rule with more neighboring cells included in
the evolution. The evolution rule is
$R(s^t_{i-2}s^t_{i-1}s^t_is^t_{i+1}s^t_{i+2})=s^{t+1}_i$ where
\begin{eqnarray}
s^{t+1}_i= \left\{
\begin{array}{cr}
1 &\; q^t_i>0\\
0 &\; q^t_i<0\\
s^t_i &\; q^t_i=0
\end{array}\right.{,}\label{eq:eleven}
\end{eqnarray}
and $q^t_i=\sum^{\mu=2}_{\mu=-2}s^t_{i+\mu}$. The $\textbf{T}$
matrix for this rule is a $32\times32$ matrix and the maximum
in-degree $k_m$ is $1414$ for $L=15$. From this example is is also
apparent that, in general for the evolution rule involves $2n$
neighboring cells, the basic $\textbf{T}$ matrix should be defined
by $2^{2n+1}\times2^{2n+1}$ matrix.

From the above discussion, one can see that with $8\times8$
$\widetilde{\textbf{T}}$ matrix, more rules can be treated within
this approach. As a consequence, such extension of $\textbf{T}$ into
larger matrix seems to provide a better description for network
analysis. Unfortunately such approach can still not be able to cover
all elementary CA. There exists rules in which the configurations of
hubstates vary according to $L$ being even or odd, such as
$\textbf{R}50$ and $\textbf{R}77$. On the other hand, there are also
cases in which $\textbf{T}$ varies with $L$ indefinitely, one of the
good examples is $\textbf{R}43$. We had analyzed $\textbf{R}43$ with
different $L$ and the result is listed in Table \ref{tab:table2}.
The number of hubstates will change with $L$, and the period of the
attractors is large, therefore it is not appropriate to calculate
$k_m$ by using $\textbf{T}$ matrix. For these situations, there is
no scale of $k_m$ with arbritrary $L$.

\begin{table} \caption{\label{tab:table2} The
configuration of hubstates in $\textbf{R}43$ with different $L$. }
\begin{ruledtabular}
\begin{tabular}{cccc}
  & &\textrm{one of the}&\textrm{number of}\\
 \textrm{$L$}&\textrm{$k_m$}&\textrm{hubstates}&\textrm{hubstates}\\
 \colrule
  10 & 5 & 0000110011 & 40 \\
  11& 8 & 00100110100 & 22 \\
  12 & 18 & 001100110100 & 4 \\
  13 & 13 & 0001100110100 & 26 \\
  14 & 13 & 00001100110100 & 54 \\
\end{tabular}
\end{ruledtabular}
\end{table}

For rules with the above irregularity special treatment is required
for each particular case. In the following we will discuss
$\textbf{R}50$ and $\textbf{R}77$ for illustration of their
complication. The maximum in-degree of $\textbf{R}50$ and
$\textbf{R}77$ can be properly described by $8\times8$
$\widetilde{\textbf{T}}$ matrices only when $L$ is even, since the
in-degree distributions will not depend with lattice length $L$.

\begin{figure}
\includegraphics[width=0.5\textwidth]{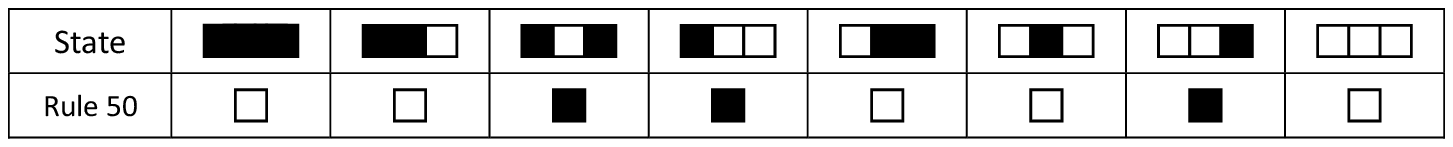}\\
\caption{\label{fig:ten} The diagram of $\textbf{R}50$.}
\end{figure}

Fig. \ref{fig:ten} shows the evolution according to $\textbf{R}50$,
and Table \ref{tab:table4} shows the relation between maximum
in-degree $k_m$ and lattice length $L$. The hubstate of
$\textbf{R}50$ is either ($0101\dots0101$) or ($1010\dots1010$) when
$L$ is even.The $k_m$ of $\textbf{R}50$ can be evaluated by
$k(S_{L=even})=1/2Tr[\widetilde{\textbf{T}}_{50}^L]$, the factor
$1/2$ is due to the fact that the $8\times8$
$\widetilde{\textbf{T}}$ matrix carries the information of all
preimages of both hubstates, namely ($0101\dots0101$) and
($1010\dots1010$). For even $\alpha$, all the preimages of
$L=\alpha+1$ can be obtained by just appending ``$0$'' or ``$1$'' at
the boundary to all the preimages of $L=\alpha$. As a consequence,
both even $L=\alpha$ and odd $L=\alpha+1$ have the same maximum
in-degree $k_m$. From these results one can see clearly that for
even $L$ $k_m$ follows scaling law with $\nu\simeq0.345$. But for
odd $L$ there is no scaling behavior for $k_m$. The same conclusion
also appears in $\textbf{R}77$ which is presentated in Table
\ref{tab:table5}.

\begin{table}
\caption{\label{tab:table4} Maximum in-degree corresponding to
lattice length $L$ in $\textbf{R}50$ by direct counting. }
\begin{ruledtabular}
\begin{tabular}{ccc|ccc}
 \textrm{$k_m$}&\textrm{even $L$}&\textrm{$\nu$}&\textrm{$k_m$}&\textrm{odd $L$}&\textrm{$\nu$}\\
 \colrule
 11 & 10&  0.3459&11 & 11 & 0.3144\\
 18 & 12 & 0.3474& 18 & 13 & 0.3207 \\
 29 & 14 & 0.3469 & 29 &15 &0.3238 \\
  47 & 16&  0.3471&  47&  17&  0.3267 \\
\end{tabular}
\end{ruledtabular}
\end{table}

\begin{figure}
\includegraphics[width=0.5\textwidth]{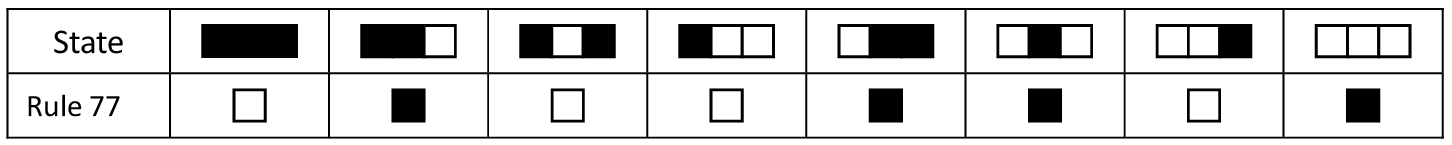}\\
\caption{\label{fig:14} The diagram of $\textbf{R}77$.}
\end{figure}

\begin{figure}
\includegraphics[width=0.25\textwidth,height=5cm]{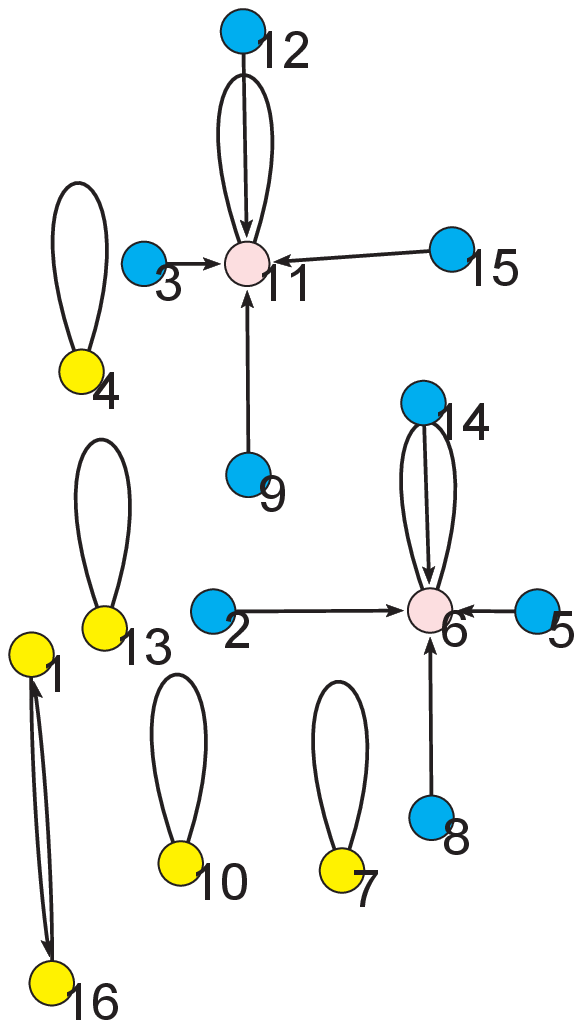}\\
\includegraphics[width=0.25\textwidth,height=5cm]{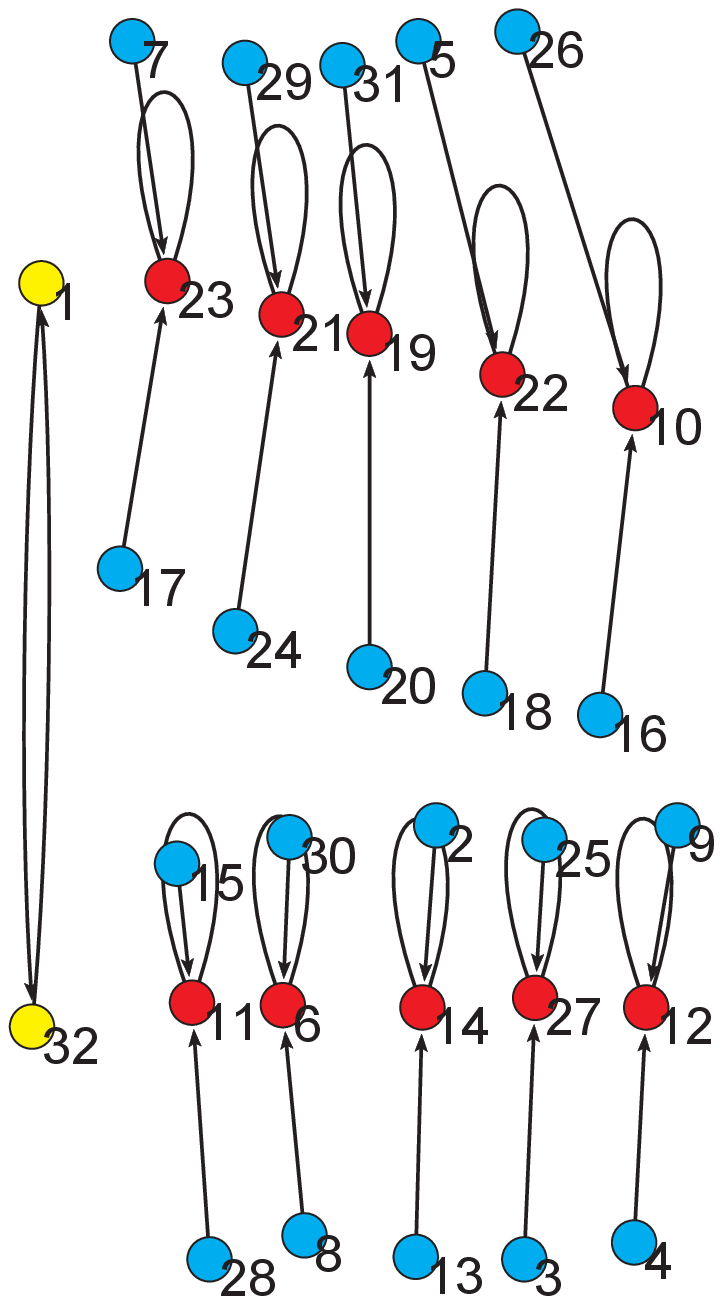}\\
\caption{\label{fig:15} The networks of $\textbf{R}77$ with (a)
$L=4$ and (b) $L=5$ which are drawn by PAJEK.}
\end{figure}

\begin{figure}
\includegraphics[width=0.5\textwidth]{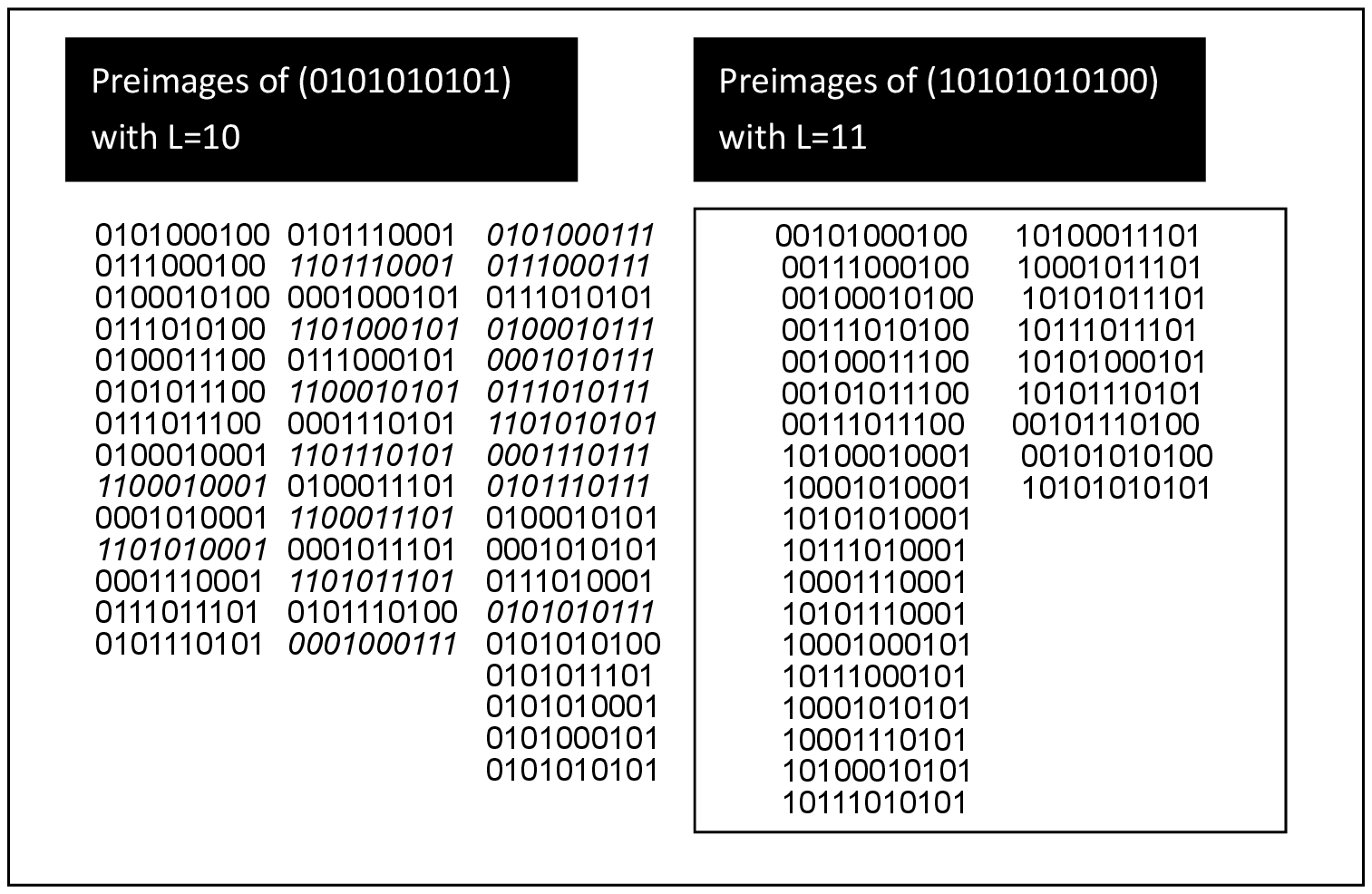}\\
\caption{\label{fig:16} The lists of pre-images of particular
hubstates in $\textbf{R}77$ with different $L$.}
\end{figure}

Rule $77$ is treated in a different way. Fig. \ref{fig:14} is the
evolution according to $\textbf{R}77$. The network configurations
are completely different for $L$ being odd or even. For illustration
it is shown in Fig. \ref{fig:15} the networks for $L=4$ and $5$.
Hence one would expect $k_m$ varies with $L$. The hubstates of the
$\textbf{R}77$ are ($0101\dots0101$) and ($1010\dots1010$) when $L$
is even, but $k_m$ of odd $L$ is quite varying, as given in Table
\ref{tab:table5}. For $\textbf{R}77$, where $L=10$ we observed that
there exist only two hubstates where $22$ hubstates are found for
$L=11$. Concentrating on one of $L=10$ hubstates, it is clear that
the hubstate of $L=11$ is just adding $1$ or $0$ to the end of
$L=10$'s hubstate. Furthermore, there are totally $46$ preimages for
$L=10$ and only $28$ for $L=11$. The $18$ italic configurations with
$L=10$ do not associate with the preimages in $L=11$. From Fig.
\ref{fig:16} it is known that only those preimages of $L=10$ without
($111$) at the boundary for can associate to the preimages of
$L=11$.

\begin{table}
\caption{\label{tab:table5} Maximum in-degree corresponding to
lattice length $L$ in $\textbf{R}77$ by direct counting. }
\begin{ruledtabular}
\begin{tabular}{ccc|ccc}
 \textrm{$k_m$}&\textrm{even $L$}&\textrm{$\nu$}&\textrm{$k_m$}&\textrm{odd $L$}&\textrm{$\nu$}\\
 \colrule
 46 & 10 & 0.552356&28 & 11 & 0.437032\\
 98 & 12 & 0.551226&  60 & 13 & 0.454376\\
   211 &14 & 0.551507&129 &15 & 0.467415\\
 453 &16 & 0.55146&  277 &17 & 0.477279\\
\end{tabular}
\end{ruledtabular}
\end{table}

To evaluate the $k_m$ with odd $L$ we follow the approach suggested
in \cite{shreim07} in obtaining arbitrary in-degree of
$\textbf{R}4$, where apart from the $\widetilde{\textbf{T}}$ matrix
itself an auxiliary matrix is constructed to resolve the difficulty.
Therefore, by this approach we propose the following formula:
\begin{equation}
k(S_{L=odd})=Tr[\widetilde{\textbf{T}}^{L-1}_{77}\times\widetilde{T}_{multiplied}].
\label{eq:12}
\end{equation}
This is due to the fact that the hubstates of odd $L$ can be viewed
as just adding ``$0$'' or ``$1$'' in the hubstates of even $L$.
According to Fig. \ref{fig:15}, we could assign the
$\widetilde{T}_{multiplied}$, which just corresponds to adding
``$0$'' in the hubstate obtained by referring to Fig. \ref{fig:16},
since the mapping at the boundary of preimages with $L=11$ has only
two possibilities:
\begin{subequations}
\begin{eqnarray}
\widetilde{T}_{multiplied}|000>=|000>. \label{eq:13a}
\\
\widetilde{T}_{multiplied}|011>=|110>. \label{eq:13b}
\end{eqnarray}
\end{subequations}
Therefore we have $\widetilde{T}_{multiplied}$ as
\begin{eqnarray}
\widetilde{T}_{multiplied}= \left(
\begin{array}{cccccccc}
1 & 0 &0 &0& 0 &0 &0 &0\\
0 &0 & 0 &0 &0& 0 &0 &0\\
0 &0 & 0 &0 &0& 0 &0 &0\\
0 &0 & 0 &0 &0& 0 &1 &0\\
0 &0 & 0 &0 &0& 0 &0 &0\\
0 &0 & 0 &0 &0& 0 &0 &0\\
0 &0 & 0 &0 &0& 0 &0 &0\\
0 &0 & 0 &0 &0& 0 &0 &0
\end{array}\right).\label{eq:14}
\end{eqnarray}
Then we can get the exact maximum in-degree of hubstate with any
lattice length $L$. This analytical results agree with direct
counting for $L\leq25$. A summary of the formulas for the maximum
in-degree of $\textbf{R}50$ and $\textbf{R}77$ are given in Table
\ref{tab:table6}.

\begin{table}
\caption{\label{tab:table6} Summary for maximum in-degree
calculation for $\textbf{R}50$ and
$\textbf{R}77$.}
\begin{ruledtabular}
\begin{tabular}{ll}
Rules & Method\\
\colrule
$\textbf{R}50$  & $k(S_{L=even})=\frac{1}{2}Tr[\widetilde{\textbf{T}}_{50}^L]$\\
&$ k(S_{L=odd})=\frac{1}{2}Tr[\widetilde{\textbf{T}}_{50}^{L-1}]$\\
 \colrule
  $\textbf{R}77$&$k(S_{L=odd})=\frac{1}{2}Tr[\widetilde{\textbf{T}}_{77}^L]$\\
  &$k(S_{L=odd})=Tr[\widetilde{\textbf{T}}^{L-1}_{77}\times\widetilde{T}_{multiplied}]$\\
\end{tabular}
 \end{ruledtabular}
\end{table}

In this section we have found more rules which could not be treated
with $4\times4$ $\textbf{T}$ matrix can now be dealt with by
$8\times8$ transfer matrix. Consequently, the maximum in-degree and
the ID can be evaluated efficiently.

\section{\label{sec:level4}CONCLUSION}
A detail discussion of the local characteristic of network generated
by elementary CA is presented in this work. First we have located
the possible source of errors in claiming the multiscaling of such
network in \cite{shreim07}. This is due to a misused of derivative
on large difference. Secondly the in-degree distributions of many
rules were calculated by direct counting method and the multiscaling
characteristic does not appear. From the studies up to $L=25$, it
suggests that the in-degree distribution of CA is $L$ independent
regardless of its classes. Furthermore all the possible largest
eigenvalues $\lambda_m$ for $4\times4$ transfer matrix $\textbf{T}$
are obtained and analysis on the correlation between
$\nu=\log_2{\lambda_m}$ and classification of CA is discussed. We
have found that there might be more rules sharing the same in-degree
distribution but corresponding to different classification according
to S. Wolfram. It has been found that there are good correlation for
class IV which possess $\nu=0.4065$, however, apart from that, the
ID of other classes do not share any universal pattern and also the
value of $\nu$ can not be used as an indicator for classification.
Therefore the in-degree distribution is not a proper characteristic
to classification which was also pointed out in \cite{shreim07}. In
section \ref{sec:level3}, the transfer matrix was extended to
$8\times8$ $\widetilde{\textbf{T}}$ matrix. This augmentation in
some sense respects the evolution of CA rule and is also more
natural to reflect the characters of hubstates' pre-images. As a
result the $8\times8$ matrix scheme was applied to several rules
which do not possess a $4\times4$ $\textbf{T}$ matrix. The
$32\times32$ $\textbf{T}$ matrix has also applied to CA beyond
neighbor evolution. The $L$ dependence of the hubstates of some
particular rules such as $\textbf{R}50$ and $\textbf{R}77$ are
discussed in detail. The calculation of their $k_m$ are also treated
in this work.

\begin{acknowledgments}
This work was supported by National Science Council of Taiwan,
NSC-97-2112-M-006-003-MY2.
\end{acknowledgments}

\end{document}